\documentclass[preprint,aps,prd,amsmath,showpacs,nofootinbib]{revtex4}
\usepackage{graphicx}

\newcommand{\beq}{\begin{equation}}
\newcommand{\eeq}{\end{equation}}
\newcommand{\bea}{\begin{eqnarray}}
\newcommand{\eea}{\end{eqnarray}}

\newcommand{\Non}{\nonumber\\}
\newcommand{\Ref}[1]{(\ref{#1})}

\begin{document}

\title{Composite vacuum Brans-Dicke wormholes}

\author{Sergey V. Sushkov$^{1,2,}$}
\email{sergey_sushkov@mail.ru} 
\affiliation{$^1$ Department of Mathematics and Department of Physics,
Kazan Federal University, Kremlevskaya str. 18, Kazan 420008, Russia\\
$^2$ Physics Department, CSU Fresno, Fresno, CA 93740-8031}

\author{Sergey M. Kozyrev}
\email{Sergey@tnpko.ru}
\affiliation{Scientific center for gravity wave studies ``Dulkyn'', Kazan, Russia}

\begin{abstract}
We construct a new static spherically symmetric configuration composed of interior and exterior Brans-Dicke vacua matched at a thin matter shell. Both vacua correspond to the same Brans-Dicke coupling parameter $\omega$, however they are described by the Brans class I solution with different sets of parameters of integration. In particular, the exterior vacuum solution has $C_{ext}(\omega)\equiv 0$. In this case the Brans class I solution for any $\omega$ reduces to the Schwarzschild one being consistent with restrictions on the post-Newtonian parameters following from recent Cassini data. The interior region possesses a strong gravitational field, and so the interior vacuum solution has $C_{int}(\omega)=-1/(\omega+2)$. In this case the Brans class I solution describes a wormhole spacetime provided $\omega$ lies in the narrow interval $-2-\frac{\sqrt{3}}{3}<\omega<-2$. The interior and exterior regions are matched at a thin shell made from an ordinary perfect fluid with positive energy density and pressure obeying the barotropic equation of state $p=k\sigma$ with $0\le k\le1$. The resulting configuration represents a composite wormhole, i.e. the thin matter shell with the Schwarzschild-like exterior region and the interior region containing the wormhole throat.
\end{abstract}


\maketitle

\section{Introduction}
Brans-Dicke theory is the famous prototype of gravitational theories alternative to Einstein's general relativity \cite{Will}. The essential feature of Brans-Dicke theory is the presence of a fundamental scalar field nonminimally coupled to curvature, and so it and its generalizations, which may include one or several
scalar fields, are generally known as scalar-tensor theories.
%\cite{Will}.
Initially the Brans-Dicke theory was developed as a modified relativistic theory of gravitation compatible with Mach's principle \cite{BD,Brans}. The current interest in scalar-tensor theories is manifold. They arise naturally as the low energy limit of many theories of quantum gravity such as superstring theories or the Kaluza-Klein theory. Moreover, Brans-Dicke and scalar-tensor theories have numerous interesting cosmological applications, which include inflationary scenarios, dark energy and dark matter models, etc (see, for example, \cite{FM}). Static
solutions in scalar-tensor theories are also of interest. In particular, Brans-Dicke wormholes have been intensively investigated \cite{AgnCam,
NanIslEva, AncGruTor, NanBhaAlaEva, BhaSar05, LobOli, BroSkvSta}.
It is worth noticing that wormhole solutions may appear in the whole ghost range of Brans-Dicke theory (i.e. for any $\omega < -3/2$; see the interesting discussion in Ref. \cite{BroSkvSta}).

The action of Brans-Dicke theory is given by\footnote{Units $8\pi G=c=1$ are used throughout the paper.}
\beq\label{action}
S =\frac12\int dx^4 \sqrt{-g}\left\{\phi R -
\omega\frac{\phi^{,\mu}\phi_{,\mu}}{\phi}\right\} +{S}_m,
\eeq
where $R$ is the scalar curvature, $\phi$ is a Brans-Dicke scalar, $\omega$ is a dimensionless coupling parameter, and ${S}_m$ is an action of ordinary matter (not including the Brans-Dicke scalar). The action \Ref{action} provides the following field equations: 
\begin{subequations} \label{BDeqs}
\bea
G_{\mu \nu } &=& \frac1\phi T_{\mu\nu}+ \frac{\omega}{\phi
^2}\phi_{,\mu}\phi_{,\nu} -\frac{\omega}{2\phi ^2} g_{\mu \nu
}\phi_{,\alpha}\phi^{,\alpha} + \frac 1\phi \phi_{;\mu;\nu} -\frac
1\phi g_{\mu \nu }\phi_{;\alpha}^{;\alpha},
\\
\phi^{;\alpha}_{;\alpha} &=&\frac{T}{2\omega+3},
\eea
\end{subequations}
where $G_{\mu \nu }=R_{\mu \nu }-\frac12 Rg_{\mu \nu }$ is the Einstein tensor, and $T=T^\mu_\mu$ is the trace of the matter energy momentum tensor $T_{\mu\nu}$. In 1962 Brans himself \cite{Brans} constructed static spherically symmetric vacuum solutions (i.e., with $T_{\mu\nu}\equiv 0$) of the Brans-Dicke field equations \Ref{BDeqs}. He found four classes of solutions, which are now known as Brans class I, II, III, and IV solutions. However, it is necessary to emphasize that solutions 
from different classes are not independent -- one can be derived from
the other by some analytical transformations \cite{BS05}. %\cite{BS05,IBN10}.
%(see the appendix).
For this reason, throughout this work we will discuss only the Brans class I solution, which is the best known spherically symmetric solution of Brans-Dicke theory.

The vacuum Brans class I solution given in isotropic coordinates reads
\begin{subequations} \label{BDclassI}
\begin{eqnarray}\label{BDmetricI}
&\displaystyle ds^2=-e^{2\nu_0} \left( \frac{1-B/r}{1+B/r}\right)
^{\frac 2 A } dt^2 +e^{2\lambda_0}\left(1+\frac Br\right)^4 \left(
\frac{1-B/r}{1+B/r}\right)^{\frac{2(A -C-1)}A } [dr^2+r^2 d\Omega^2],& \\
&\displaystyle  \phi(r) = \phi _0\left( \frac{1-B/r}{1+B/r}\right)
^{\frac C A },&
\end{eqnarray}
\end{subequations}
where $d\Omega^2=d\theta^2+\sin^2\theta d\varphi^2$ is the linear element of the unit sphere, and the radial coordinate $r$ satisfies the condition $r>B$ in order to provide an analyticity of the solution. Generally, the solution depends on five free parameters: $\phi_0$, $\nu_0$,
$\lambda_0$, $B$, and $C$. The sixth parameter $A$ is not free, it
obeys the following constraint condition:
\beq \label{A}
A=\left[(C+1)^2-C\left(1-\textstyle\frac12\omega
C\right)\right]^{1/2}>0.
\eeq
In Brans-Dicke theory the metric \Ref{BDmetricI} represents an exterior gravitational field of some spherical distribution of matter. Far from a source of gravity, i.e., in the limit $r\to\infty$, it takes the form
\beq\label{asymp}
ds^2=-e^{2\nu_0}\left(1-\frac{2M}{r}+O(r^{-2})\right)
dt^2+e^{2\lambda_0}\left(1+\frac{2\gamma
M}{r}+O(r^{-2})\right)[dr^2+r^2 d\Omega^2],
\eeq
where $M=2B/A$ is an asymptotic mass measured by a distant observer, and $\gamma=1+C$ is the post-Newtonian parameter. Because of asymptotic flatness one should set $\nu_0=\lambda_0=0$. The value of $\gamma$ can be estimated from the recent conjunction experiment with Cassini spacecraft as $|\gamma-1|\le 2.3\times 10^{-5}$ \cite{Cassini1,Cassini2}. Hence, one get
\beq\label{C}
|C|\le 2.3\times 10^{-5}.
\eeq
Note that, formally, the parameter $C$ does not depend on $\omega$, and so one may directly set $C=0$ in Eqs. \Ref{BDclassI} and \Ref{A}. As the result, one find $A=1$ and
\begin{subequations} \label{Schwarz}
\begin{eqnarray}\label{Smetric}
&\displaystyle ds^2=- \left(
\frac{1-M/2r}{1+M/2r}\right) ^{2 } dt^2 + \left( 1+\frac{M}{2r}\right)^4
[dr^2+r^2 d\Omega^2],& \\
& \phi(r)\equiv\phi_0=const, &
\end{eqnarray}
\end{subequations}
where Eq. \Ref{Smetric} is nothing but the Schwarzschild metric (in isotropic coordinates). It is obvious that the Schwarzschild solution is perfectly consistent with observational data. However, one should remember that any exterior vacuum Brans-Dicke solution have to be matched to some interior one. Supposing that the interior Brans-Dicke solution corresponds to some reasonable spherical distribution of matter, one can get on the basis of a post-Newtonian weak field approximation the following relationship \cite{Weinberg}:
\beq\label{Comega}
C(\omega)=-\frac{1}{\omega+2}.
\eeq
Now the limiting (Schwarzschild) case $C(\omega)\to 0$, $A(\omega)\to 1$ is only possible under the limit $|\omega|\to\infty$.\footnote{Here it should be mentioned that Brans-Dicke theory (and its dynamic generalization) in the limit $|\omega|\to\infty$ reduces to general relativity \cite{Weinberg} (see, also, Refs. \cite{traceless1,traceless2,traceless3,traceless4,traceless5} where the specific case of a traceless energy momentum tensor is discussed).} Using Eqs. \Ref{C} and \Ref{Comega} one can find the lower boundary for $\omega$: $|\omega|\ge 5\times 10^4$. Thus, the consideration based on the post-Newtonian weak field approximation leads to the conclusion that the Brans-Dicke theory can be consistent with the (local) observations only if $\omega$ is very large.

On the other hand there is no reason for the relationship \Ref{Comega} to hold in the presence of compact objects with strong gravitational fields. For example, in the context of gravitational collapse in Brans-Dicke theory, Matsuda \cite{Matsuda} had considered $C(\omega)\propto -\omega^{-1/2}$. The other examples of essentially relativistic objects possessing strong gravitational fields are represented by wormholes. Vacuum Brans-Dicke wormholes with various $C(\omega)$ were discussed in the literature. Namely, in Ref. \cite{NanBhaAlaEva} the case $C(\omega) = -q\omega^{-1/2}$ with $q < 0$ had been considered . Also, Lobo and Oliveira \cite{LobOli} discussed two models: $C(\omega)=(\omega^2+\omega_0^2)^{-1}$, and $C(\omega)=\lambda\exp(-\omega^2/2)$.

In this paper we will accept a more general conjecture. Namely, we will suppose that the form of $C(\omega)$ can be, in principle, different in various spacetime regions. In other words this means that various spacetime regions can possess different Brans-Dicke vacua. To justify this supposition one can speculate that Brans-Dicke vacuum states are forming due to phase transitions in some generalized theory, and the vacuum state formation is depending on local values of the gravitational field. As the result one will obtain `bubbles' of different Brans-Dicke vacua  divided by 'walls'.

Applying the conjecture about a variety of Brans-Dicke vacua, we will consider a simple static spherically symmetric configuration composing of two different vacua.

%%%%%%%%%%%%%%%%%%%%%%%%%%%%%%%%%%%%%%%%%%%%%%%%%%%%%%%%%
\section{Composite vacuum solution}
%%%%%%%%%%%%%%%%%%%%%%%%%%%%%%%%%%%%%%%%%%%%%%%%%%%%%%%%%
Let us consider a static spherically symmetric configuration composed of two Brans-Dicke vacua.
The spacetime metric in both interior and exterior regions is given in isotropic coordinates as follows
\beq\label{metric}
ds^2=-e^{2\nu(r)}dt^2+e^{2\lambda(r)}[dr^2+r^2d\Omega^2],
\eeq
so that $x^\gamma=(t,r,\theta,\phi).$ We will assume that the interior
is described by the vacuum Brans class I solution:
\begin{subequations} \label{interior}
\begin{eqnarray}
e^{\nu_{int}(r)} &=& e^{\nu_{0}} \left( \frac{1- B/r}{1+
B/r} \right) ^{\frac{1}{A} }, \\
e^{\lambda_{int}(r)} &=& e^{\lambda_{0}}\left( 1+\frac{B}{r}\right) ^2\left(
\frac{1-B/r}{1+ B/r}\right) ^{\frac{A -C-1}{A} }, \label{glambda}\\
\phi_{int}(r) &=& \phi_{0}\left( \frac{1- B/r}{1+ B/r}\right)
^{\frac{C}{A} },
\end{eqnarray}
\end{subequations}
where $\phi_{0}$, $\nu_{0}$, $\lambda_{0}$, $B$, and $C$ are free (still
undefined) parameters, and $A$ is given by Eq. \Ref{A}. Note that the radial coordinate $r$ runs monotonically from $B$ to $a$, where $a>B$ is a boundary of the interior region.

As was already mentioned, an exterior region of some spherical gravitating configuration can be also described by the Brans class I solution provided that $C$ fulfils the constraint \Ref{C}. Assuming $C_{ext}(\omega)\equiv 0$, we obtain the exterior Schwarzschild solution:
\begin{subequations} \label{exterior}
\begin{eqnarray}
e^{\nu_{ext}(r)} &=&
\frac{1-M/2r}{1+M/2r},\\
e^{\lambda_{ext}(r)}  &=& \left( 1+\frac{M}{2r}\right) ^2, \\
\phi_{ext}(r) &=& 1,
\end{eqnarray}
\end{subequations}
where $M=2B_{ext}$ is the Schwarzschild mass. Note that we have put $\lambda_{0,ext}=\nu_{0,ext}=0$ in order to provide the asymptotic flatness. Also, without loss of generality, we put $\phi_{0,ext}=1$. The radial coordinate $r$ within the exterior region runs from $a$ to infinity. We will suppose that $a>M/2$; this guarantees that the exterior region does not contain the event horizon.

So, $r$ is the global radial coordinate monotonically running from $B$ to $a$ in the interior region, and from $a$ to infinity in the exterior one. The surface $\Sigma: r=a$ is a thin shell where the interior and exterior solutions, \Ref{interior} and \Ref{exterior}, should be matched. Since Eqs. \Ref{interior} and \Ref{exterior} are the vacuum Brans-Dicke solutions, we should conclude that all ordinary matter (excluding the Brans-Dicke scalar) is concentrated at the thin shell $\Sigma$.

Here it is worth noticing that thin-shell Brans-Dicke wormholes were studied in the literature \cite{thin1,thin2}. The models considered in Refs. \cite{thin1,thin2} were constructed by the cut-and-paste method.\footnote{The first examples of thin-shell wormholes have been given by Visser \cite{Vis1,Vis2}.} Though our construction seems to be similar to cut-and-paste wormhole configurations, this similarity has only a formal character. Actually, the essence of the method is the following: One takes two the same copies of spacetime manifolds with appropriate asymptotics, cuts and casts away 'useless' regions of spacetimes (containing horizons, singularities, etc.), and pastes remaining regions. As the result, one obtains a geodesically complete wormhole spacetime with given asymptotics (Schwarzschild, Reissner-Nordstrom, Brans-Dicke, etc.) and a throat being a thin shell of exotic matter violating the null energy condition. In our case, we have initially a thin shell made from \textit{ordinary} matter, and then we look for appropriate interior and exterior Brans-Dicke solutions matched at the shell. Note that this approach is similar to the problem of a thin shell in general relativity (see Ref. \cite{thinshellingr}). However, the distinction is that the Birkhoff theorem is not valid in Brans-Dicke theory, and so both the interior and exterior Brans-Dicke vacua are not unique.

To analyze a thin-shell configuration we will follow the standard Darmois-Israel formalism \cite{DI}, also known as the junction condition formalism. The shell $\Sigma$ is a synchronous timelike hypersurface with intrinsic coordinates $\xi^{i}=(\tau,\theta,\varphi)$. The coordinate $\tau$ is the proper time on the shell. Generally, a position of $\Sigma$ can be a function of the proper time. However, hereafter we will assume $a(\tau)\equiv a=const$. Note that the metric (first fundamental form) and the scalar field should be continuous on $\Sigma$:
\beq\label{matchingcond}
\lambda_{int}(a)=\lambda_{ext}(a), \quad
\nu_{int}(a)=\nu_{ext}(a), \quad
\phi_{int}(a)=\phi_{ext}(a)\equiv 1.
\eeq
Substituting Eqs. \Ref{interior} and \Ref{exterior} into \Ref{matchingcond} gives
\begin{subequations}\label{coefs}
\bea
e^{\nu_0}&=&\left(\frac{1-M/2a}{1+M/2a}\right)\left(\frac{1+B/a}{1-B/a}\right)^{\frac{1}{A}},\\
e^{\lambda_0}&=&\left(1+\frac{M}{2a}\right)^2 \left(1+\frac{B}{a}\right)^{-2} \left(\frac{1+B/a}{1-B/a}\right)^{\frac{A-C-1}{A}},\\
\phi_0&=&\left(\frac{1+B/a}{1-B/a}\right)^{\frac{C}{A}}.
\eea
\end{subequations}
At the same time, derivatives of the metric and scalar field can be discontinuous. The discontinuity of the metric is usually described in terms of a jump of the extrinsic curvature $K_{ij}$. The extrinsic curvature (second fundamental form) associated with a hypersurface $\Sigma: F(x)=0$ is given by
\begin{equation}\label{second_form}
K_{ij}= \left. -n_{\gamma}\left( \frac{\partial^2
x^{\gamma}}{\partial\xi^{i}\partial\xi^{j}}+\Gamma^\gamma_{\alpha\beta}
\frac{\partial x^{\alpha}}{\partial\xi^{i}} \frac{\partial
x^{\beta}}{\partial\xi^{j}}\right)\right|_{\Sigma},
\end{equation}
where $n_\gamma$ is the unit normal ($n^\gamma n_\gamma=1$) to $\Sigma$:
\beq
n_\gamma=\left|g^{\alpha\beta}\frac{\partial F}{\partial
x^\alpha} \frac{\partial F}{\partial x^\beta}\right|^{-1/2}
\frac{\partial F}{\partial x^\gamma}.
\eeq
The junction conditions in Brans-Dicke theory (generalized Darmois-Israel conditions) can be obtained by projecting on $\Sigma$ the field equations \Ref{BDeqs} \cite{surfaceeqn}:
\beq\label{surfK}
-[K_j^i]+[K]\delta_j^i=\frac{8\pi}{\phi}\left(S_j^i-\frac{S}{3+2\omega}\,\delta_j^i\right),
\eeq
\beq\label{surfphi}
[\phi_{,n}]=\frac{8\pi S}{3+2\omega},
\eeq
where the notation $[Z]=Z_{ext}|_{\Sigma}-Z_{int}|_{\Sigma}$ stands for the jump of a given quantity $Z$ across the hypersurface $\Sigma$, $n$ labels the coordinate normal to this surface and $S_{ij}$ is the energy-momentum tensor of matter on the shell located at $\Sigma$. The quantities $K$ and $S$ are the traces of $K^i_j$ and $S^i_j$ respectively. Note that Eq. \Ref{surfK} is equivalent to
\beq\label{surfS}
S^i_j=\frac{\phi}{8\pi}\left(\frac{\omega+1}{\omega}[K]\delta^i_j-[K^i_j]\right).
\eeq

The jump of the components of the extrinsic curvature associated with two sides of the hypersurface $F(x)=r-a=0$ in the spacetime with the metric \Ref{metric} can be found as
\beq
[K^\tau_\tau]=[\nu']e^{-\lambda(a)},\quad
[K^\theta_\theta]=[K^\varphi_\varphi]= [\lambda']e^{-\lambda(a)}.
\eeq
The surface stress-energy tensor of a perfect fluid is given by
\begin{equation} \label{Sperfluid}
S^i_j=
\left(
\begin{array}{ccc}
-\sigma & 0 & \phantom{.}0 \\
0 & p & \phantom{.}0\\
0 & 0 & \phantom{.}p
\end{array}
\right),
\end{equation}
where $\sigma$ and $p$ are the surface energy density and the surface pressure, respectively. Now, Eq. \Ref{surfS} yields
\bea\label{eqs}
\sigma&=&-\frac{\phi(a) e^{-\lambda(a)}}{8\pi\omega} \left([\nu']+2(\omega+1)[\lambda']\right),
\\ \label{eqp}
p&=&\frac{\phi(a) e^{-\lambda(a)}}{8\pi\omega} \left((\omega+1)[\nu']+(\omega+2)[\lambda']\right).
\eea
where $[\nu']=\nu'_{ext}(a)-\nu'_{int}(a)$ and $[\lambda']=\lambda'_{ext}(a)-\lambda'_{int}(a)$. The obtained relations express the surface energy density $\sigma$ and the surface pressure $p$ in terms of jumps of first derivatives of the metric functions. Substituting the expressions  \Ref{interior}-\Ref{coefs} for metric coefficients into \Ref{eqs} and \Ref{eqp} we find
\bea
\label{sigma}
\sigma&=&\frac{1}{4\pi a\omega(1+\mu)^2}\left[\frac{\mu ( 2\omega(1-\mu)-2\mu+1 )}{1-\mu^2} -\frac{\beta ( 2\omega(1-\beta A+C) -2\beta A+2C+1 )}{A(1-\beta^2)} \right],\\
\label{p}
p&=&\frac{1}{4\pi a\omega(1+\mu)^2}\left[\frac{\mu(\omega\mu+2\mu-1)}{1-\mu^2} -\frac{\beta(\omega(\beta A-C) +2\beta A-2C-1)}{A(1-\beta^2)} \right],
\eea
where $\mu=M/2a$ and $\beta=B/a$ are convenient dimensionless values such that $\mu<1$ and $\beta<1$.

%%%%%%%%%%%%%%%%%%%%%%%%%%%%%%%%%%%%%%%%%%%%%%%%%%%%%%
\section{Matter on the thin shell}
%%%%%%%%%%%%%%%%%%%%%%%%%%%%%%%%%%%%%%%%%%%%%%%%%%%%%%
Resulting expressions \Ref{sigma} and \Ref{p} give the energy density and the pressure of matter filling the thin shell $\Sigma$ in terms of parameters of the model: the coupling parameter $\omega$, the shall radius $a$, the exterior vacuum parameter $\mu$ (dimensionless Schwarzschild mass), and the interior vacuum parameters $\beta$ and $C$. Note that both $\sigma$ and $p$ are proportional to $a^{-1}$, and so without lost of generality we can make rescaling $\sigma\to\sigma a^{-1}$ and $p\to p a^{-1}$, or, equivalently, just put $a=1$. To proceed further investigations we will fix the specific form of $C(\omega)$ given by Eq. \Ref{Comega}, so that $C(\omega)=-1/(\omega+2)$. In this case, Eq. \Ref{A} yields
\beq\label{Aomega}
A(\omega)=\sqrt{\frac{2\omega+3}{2\omega+4}}.
\eeq
Note that the expression under the square root in Eq. \Ref{Aomega} is positive provided $\omega<-2$ or $\omega>-3/2$. Hereafter we will restrict our consideration to the case $\omega<-2$, since only this case includes Brans-Dicke wormhole configurations.

\begin{figure}
\begin{center}
\includegraphics[width=7cm]{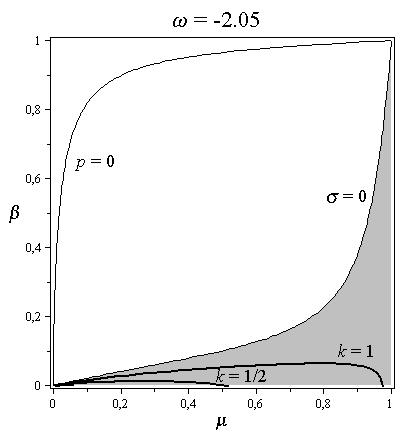} \includegraphics[width=7cm]{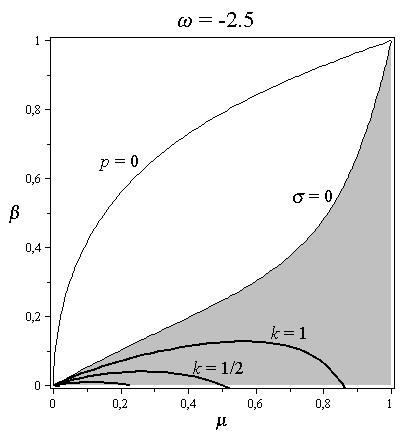}
\includegraphics[width=7cm]{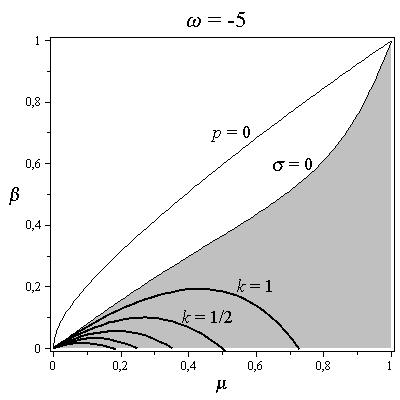} %\includegraphics[width=6cm]{A-15.jpg}
\end{center}
\caption{Contour plots for $\sigma(\mu,\beta,\omega_0)$ and $p(\mu,\beta,\omega_0)$ for $\omega_0=-2.05$, $-2.5$, and $-5$. Thin solid curves denote lines of zero level: $\sigma(\mu,\beta,\omega_0)=0$ (lower line) and $p(\mu,\beta,\omega_0)=0$ (upper line). A corresponding value is positive in the region below the zero level line, and so both $\sigma>0$ and $p>0$ in the shadowed region. Thick solid curves denote lines given by the equation of state $p(\mu,\beta,\omega_0)-k\sigma(\mu,\beta,\omega_0)=0$, where $k=1, \frac12, \frac13, \frac14, \frac15$ from top to down.
\label{fig1}}
\end{figure}

Substituting given $C(\omega)$ and $A(\omega)$ into \Ref{sigma} and \Ref{p} yields
\begin{subequations}\label{sp}
\bea
\sigma(\mu,\beta,\omega)&=&\frac{1}{4\pi\omega(1+\mu)^2}\Bigg[\frac{\mu ( 2\omega(1-\mu)-2\mu+1 )}{1-\mu^2}\Non
&&\phantom{\frac{1}{4\pi\omega(1+\mu)^2}[}
-\left(\frac{2\omega+4}{2\omega+3}\right)^{1/2}\frac{\beta ( 2\omega(1-\tilde\beta(\omega)) -2\tilde\beta(\omega)+1 )}{1-\beta^2} \Bigg],\\
p(\mu,\beta,\omega)&=&\frac{1}{4\pi\omega(1+\mu)^2}\Bigg[\frac{\mu(\omega\mu+2\mu-1)}{1-\mu^2}\Non &&\phantom{\frac{1}{4\pi\omega(1+\mu)^2}[} -\left(\frac{2\omega+4}{2\omega+3}\right)^{1/2}\frac{\beta(\omega\tilde\beta(\omega) +2\tilde\beta(\omega)-1)}{1-\beta^2} \Bigg],
\eea
\end{subequations}
where
$$
\tilde\beta(\omega)=\beta \sqrt{\frac{2\omega+3}{2\omega+4}}+\frac{1}{\omega+2}.
$$
Fixing a particular value of $\omega=\omega_0$, we obtain $\sigma$ and $p$ as functions of $\mu$ and $\beta$.
%For example, in case $\omega=-13/6$ we get
%\bea
%\sigma(\mu,\beta)&=& \frac{(7-10\mu)\beta^2-26(1-\mu^2)\beta+10\mu-7\mu^2} {26\pi (1+\mu)^2(1-\mu^2)(1-\beta^2)},\Non
%p(\mu,\beta)&=&\frac{-(1+6\mu)\beta^2+6\mu+\mu^2} {52\pi (1+\mu)^2(1-\mu^2)(1-\beta^2)}.\non
%\eea
In Fig. \ref{fig1} we present a series of contour plots for $\sigma(\mu,\beta,\omega_0)$ and $p(\mu,\beta,\omega_0)$ on the $(\mu,\beta)$-plane for various values of $\omega_0$. It is worth noticing that the plots demonstrate that for all $\omega<-2$ there are domains such that both $\sigma$ and $p$ are positive.

Additionally, any reasonable model of matter should include an equation of state $p=p(\sigma)$ imposing some relation between the energy density $\sigma$ and the pressure $p$. Hereafter we will consider the barotropic equation of state $p=k\sigma$. Note that the equation-of-state parameter $k$ is non-negative, $k\ge0$, for ordinary matter with positive energy density and non-negative pressure. Moreover, the condition $k\le1$ guarantees that the speed of sound in matter medium does not exceed the speed of light. In particular, $k=0$ for the dust, $k=1/3$ for the radiation, $k=1$ for the stiff matter. By using Eqs. \Ref{sp}, the equation of state can be rewritten as
\beq\label{EoS}
p(\mu,\beta,\omega)-k\sigma(\mu,\beta,\omega)=0.
\eeq
%For example, in case $\omega=-13/6$ Eq. \Ref{EoS} yields
%$$
%(1+6\mu+2k(7-10\mu))\beta^2-52k(1-\mu^2)\beta+2k(10\mu-7\mu^2)-6\mu-\mu^2=0.
%$$
For given $\omega$ and $k$ this equation provides an additional relation between $\mu$ and $\beta$ which can be graphically represented as a some curve on the $(\mu,\beta)$-plane. In Fig. \ref{fig1} we show such curves given for different values of $\omega$ and $k$.
%; the more detailed plot given for the specific value $\omega=2.5$ is shown in Fig.~\ref{fig2}. 
The graphical representation illustrates that for any $\omega<-2$ and $k>k_{min}(\omega)$, there exist a domain of $(\mu,\beta)$, where $\mu\in(0,\mu_{max}(\omega,k))$ and $\beta\in(0,\beta_{max}(\omega,k))$, such that the equation of state \Ref{EoS} holds. Note that a boundary value $k_{min}$ depends on $\omega$, and $\mu_{max}$ and $\beta_{max}$ depend both on $\omega$ and $k$. Note also that since $k_{min}(\omega)>0$, then $p>0$, and so matter filling the thin shell is not the dust with zero pressure.

%\begin{figure}
%\begin{center}
%\includegraphics[width=6cm]{eos.jpg}
%\end{center}
%\caption{Contour plots for $\sigma(\mu,\beta)$ and $p(\mu,\beta)$ in the case $C(\omega)=-1/(\omega+2)$. Solid curves mark lines of zero level: $\sigma=0$ (lower line) and $p=0$ (upper line). The corresponding value, $\sigma$ or $p$, is positive in the region under the zero level line. In the shadowed region both $\sigma>0$ and $p>0$.
%\label{fig2}}
%\end{figure}

Finally, we may conclude that the thin shell dividing two static spherically symmetric regions with different Brans-Dicke vacua can be made from ordinary matter. In particular, it can be the perfect fluid with the barotropic equation of state.

%%%%%%%%%%%%%%%%%%%%%%%%%%%%%%%%%%%%%%%%%%%%%%%%%%%%%%
\section{Composite configuration with a wormhole}
%%%%%%%%%%%%%%%%%%%%%%%%%%%%%%%%%%%%%%%%%%%%%%%%%%%%%%
In previous sections we have constructed the static spherically symmetric configuration composed of two Brans-Dicke vacua and demonstrated that the thin shell dividing the regions with different vacua can be made from ordinary matter. In this section we will discuss the problem: Under which conditions does the composite configuration represent a wormhole?

The exterior region of the composite configuration is described by the Schwarzschild metric and does not contain any wormholes. Let us consider the interior region. The interior metric \Ref{interior} has an explicit singular behavior at $r=B$.  To determine either it is a real or fictitious (coordinate) singularity, we should explore a behavior of curvature invariants. For example, the scalar curvature calculated in the metric \Ref{interior} reads
\beq\label{R}
R = -\frac{8 e^{-2\lambda_0}r^4
B^2(A^2-C^2-C-1)}{A^2(r-B)^{2(2A-C-1)/A} (r+B)^{2(2A+C+1)/A}}.
\eeq
It is obvious that $R$ becomes to be singular in points where the
denominator of Eq. \Ref{R} is equal to zero. In particular, if the
power of the term $(r-B)^{2(2A-C-1)/A}$ is positive then $r=B$ is
a naked singularity. And vice versa, the scalar curvature $R$
remains being regular at $r=B$ provided the power of
$(r-B)^{2(2A-C-1)/A}$ is negative. Substituting $C(\omega)$ and $A(\omega)$ given by Eqs. \Ref{Comega} and \Ref{Aomega} into the
inequality $2(2A-C-1)/A<0$ we obtain
\beq
2-\frac{\omega+1}{\omega+2}\sqrt{\frac{2\omega+4}{2\omega+3}}<0.
\eeq
The last inequality is fulfilled in a narrow interval
\beq\label{intom}
-2-\frac{\sqrt{3}}{3}<\omega<-2.
\eeq
Hence the scalar curvature $R$ is regular at $r=B$ if and only if $\omega$ takes its value within the interval \Ref{intom}. Moreover, since $R\propto (r-B)^{2|2A -C-1|/A}$, it is equal to zero at $r=B$. Note that in this case the metric function $e^{2\lambda_{int}(r)}$ given by Eq. \Ref{glambda} tends to infinity as $r\to B$, and hence $r=B$ is a flat spacial infinity.

Finally, the composite vacuum configuration with $-2-\frac{\sqrt{3}}{3}<\omega<-2$ is regular in the range $r\in(B,\infty)$, does not contain horizons in this range, and is asymptotically flat both as $r\to B$ and $r\to\infty$. Therefore, we can conclude that such the configuration is nothing but a wormhole.

Let us determine a position of the wormhole throat. It corresponds to a sphere $r=r_{th}$ with the radius $r_{th}$ providing a global minimum of the function $r^2e^{2\lambda_{int}(r)}$ (this guarantees the minimality of area of the sphere). The value $r_{th}$ is called a throat radius. Taking into account Eq. \Ref{glambda}, we find
\beq\label{throat}
r_{th}=B\left[\frac{C+1}{A}+\sqrt{\left(\frac{C+1}{A}\right)^2-1}\right].
\eeq
%where
%\beq
%\gamma(\omega)=\frac{C(\omega)+1}{A(\omega)}+\sqrt{\left(\frac{C(\omega)+1}{A(\omega)}\right)^2-1}.
%\eeq
Substituting Eqs. \Ref{Comega} and \Ref{Aomega} for $C(\omega)$ and $A(\omega)$ into the last expression yields
\beq
r_{th}=B\gamma(\omega),
\eeq 
where
$$
\gamma(\omega)=\frac{\omega+1}{\omega+2}\sqrt{\frac{2\omega+4}{2\omega+3}}+ \sqrt{\frac{-3\omega-4}{(2\omega+3)(\omega+2)}}.
$$
One can easily check that $\gamma(\omega)>1$ for any $\omega<-2$, and hence $r_{th}>B$.

Since $B=a\beta$, we have $r_{th}=a\beta\gamma(\omega)$. As was shown in the previous section, if the thin shell is made from the perfect fluid with the barotropic equation of state $p=k\sigma$, then $\beta<\beta_{max}(\omega,k)$. A numerical analysis shows that $\beta_{max}(\omega,k)\gamma(\omega)<1$ for any $\omega<-2$ and $k>k_{min}(\omega)$, and so  $r_{th}<a$. Therefore, we can conclude that the wormhole throat $r=r_{th}$ is
situated within the interior region $r\in(B,a)$ of composite vacuum configuration.

%%%%%%%%%%%%%%%%%%%%%%%%%%%%%%%%%%%%%%%%%%%%%%%%%%%%%%%%%
\section{Summary}
%%%%%%%%%%%%%%%%%%%%%%%%%%%%%%%%%%%%%%%%%%%%%%%%%%%%%%%%%
In this paper we have constructed a new static spherically symmetric configuration composed 
of interior and exterior Brans-Dicke vacua divided by thin matter shell.
Both vacua correspond to the same Brans-Dicke coupling parameter $\omega$, however they are described by the Brans class I solution \Ref{BDclassI} with different sets of parameters of integration. In particular, the exterior vacuum solution has $C_{ext}(\omega)\equiv 0$. In this case the Brans class I solution with any $\omega$ just reduces to the Schwarzschild one being consistent with restrictions on the post-Newtonian parameters following from recent Cassini data. The interior region possesses a strong gravitational field, and so, generally, $C_{int}(\omega)\not=0$. In particular, we have used a specific choice $C(\omega)=-1/(\omega+2)$. In this case the Brans class I solution describes a wormhole provided $\omega$ lies in the narrow interval $-2-\frac{\sqrt{3}}{3}<\omega<-2$. The interior and exterior regions are matched at a thin shell made from ordinary matter with positive energy density and pressure. We have studied in detail the shell made from a perfect fluid with the barotropic equation of state $p=k\sigma$ with $0\le k\le1$. The resulting configuration represents a composite wormhole, i.e. the thin matter shell with the Schwarzschild-like exterior region and the interior region containing the wormhole throat.

%Let us emphasize some features of composite wormholes. First of all, it should be noted that 
An interesting feature of composite wormholes is that the strong-field interior region containing all exotic ghost-like matter is hidden behind the matching surface, whereas the weak-field region out of it possesses the usual Schwarzschild vacuum. Such the configuration is similar to the model of trapped-ghost wormholes \cite{BroSus}. Note that in both models wormholes are twice asymptotically flat. However, in the trapped-ghost wormhole model the ghost is hidden in some restricted region around the throat, whereas in the composite wormhole model the ghost-like Brans-Dicke scalar occupies the ``half'' of wormhole spacetime behind the matching surface. Anyway, in the composite wormhole configuration a "ghost" is hidden in the strong-field interior region, which may in principle explain why no ghosts are observed under usual conditions.

%----------------------------------------------------------------
\section*{Acknowledgments}
%----------------------------------------------------------------
The work was supported in part by the Russian Foundation for Basic Research grants No. 11-02-01162. Also S.S. appreciates Douglas Singleton and California State University Fresno for hospitality during the Fulbright scholarship visit.

\end{document}